\documentclass[twocolumn]{revtex4-1}

\usepackage{latexsym}
\usepackage{amssymb}
\usepackage{amsmath}
\usepackage{amsthm}
\usepackage{amsfonts}
\usepackage{ifthen}
\usepackage{revsymb}
\usepackage{yfonts}
\usepackage{graphicx}
\usepackage{algpseudocode}
\usepackage{bbm}
\usepackage{multirow}
\usepackage{gensymb}
\usepackage{epstopdf}
\usepackage{color}
\usepackage{afterpage}
\usepackage{verbatim}
\usepackage{subdepth}
\usepackage{soul}
\usepackage{lineno, blindtext}

\usepackage[colorlinks = true]{hyperref}

\usepackage{xcolor}
\definecolor{darkred}  {rgb}{0.5,0,0}
\definecolor{darkblue} {rgb}{0,0,0.5}
\definecolor{darkgreen}{rgb}{0,0.5,0}

\hypersetup{
  urlcolor   = blue,         
  linkcolor  = darkblue,     
  citecolor  = darkgreen,    
  filecolor  = darkred       
}

\newcommand{\be}{\begin{equation}}
\newcommand{\ee}{\end{equation}}
\newcommand{\bq}{\begin{eqnarray}}
\newcommand{\eq}{\end{eqnarray}}
\newcommand{\bea}{\begin{eqnarray}}
\newcommand{\eea}{\end{eqnarray}}
\newcommand{\ba}{\begin{align}}
\newcommand{\ea}{\end{align}}

\newcommand{\ket}[1]{ | \, #1 \rangle}

\newcommand{\no}{\nonumber\\}

\definecolor{mygray}{gray}{0.6}

\definecolor{mygray}{gray}{0.9}

\newcommand{\beginsupplement}{%
        \setcounter{table}{0}
        \renewcommand{\thetable}{S\arabic{table}}%
        \setcounter{figure}{0}
        \renewcommand{\thefigure}{S\arabic{figure}}%
     }

\begin{document}

\title{Extending the computational reach of a noisy superconducting quantum processor}

\author{Abhinav Kandala, Kristan Temme, Antonio D. C{\'o}rcoles, Antonio Mezzacapo, Jerry M. Chow, Jay M. Gambetta}

\affiliation{IBM T.J.  Watson  Research  Center,  Yorktown  Heights,  NY 10598,  USA}

\date{\today}

\maketitle
\noindent

\textbf{Quantum computation, a completely different paradigm of computing, benefits from theoretically proven speed-ups for certain problems and opens up the possibility of exactly studying the properties of quantum systems~\cite{Feynman}. Yet, because of the inherent fragile nature of the physical computing elements, qubits, achieving quantum advantages over classical computation requires extremely low error rates for qubit operations as well as a significant overhead of physical qubits, in order to realize fault-tolerance via quantum error correction~\cite{Shor1995,Steane1996}. However, recent theoretical work ~\cite{Mitigation3,Mitigation2} has shown that the accuracy of computation based off expectation values of quantum observables can be enhanced through an extrapolation of results from a collection of varying noisy experiments. Here, we demonstrate this error mitigation protocol on a superconducting quantum processor, enhancing its computational capability, with no additional hardware modifications. We apply the protocol to mitigate errors on canonical single- and two-qubit experiments and then extend its application to the variational optimization ~\cite{yung2013transistor,Peruzzo13,farhi2014quantum} of Hamiltonians for quantum chemistry and magnetism~\cite{Kandala2017}. We effectively demonstrate that the suppression of incoherent errors helps unearth otherwise inaccessible accuracies to the variational solutions using our noisy processor. These results demonstrate that error mitigation techniques will be critical to significantly enhance the capabilities of near-term quantum computing hardware.}

Quantum computation can be extended indefinitely if decoherence and inaccuracies in the implementation of gates can be brought below an error-correction threshold ~\cite{Shor1995,Steane1996}. However, the resource requirements for a fully-fault tolerant architecture lie beyond the scope of near-term quantum hardware~\cite{Preskill2018}. In the absence of quantum error correction, the dominant sources of noise in current hardware are unitary gate errors and decoherence, both of which set a limit on the size of the computation that can be carried out.  In this context, hybrid-quantum algorithms ~\cite{Peruzzo13,farhi2014quantum,mcclean2016theory} with short-depth quantum circuits have been designed to perform computations within the available coherence window, while also demonstrating some robustness to coherent unitary errors~\cite{OMalley16,Kandala2017}. However, even when restricting to short depth circuits, the effect of decoherence already becomes evident for small experiments ~\cite{Kandala2017}. The recently proposed zero-noise extrapolation method~\cite{Mitigation3,Mitigation2,Endo2017arxiv} presents a route to mitigating incoherent errors and significantly improving the accuracy of the computation. It is important to note that, unlike quantum error-correction this technique does not allow for an indefinite extension of the computation time, and only provides corrections to expectation values, without correcting for the full statistical behavior. However, since it does not require any additional quantum resources, the technique is extremely well suited for practical implementations with near-term hardware.

\begin{figure*}
\includegraphics[width=6.5in]{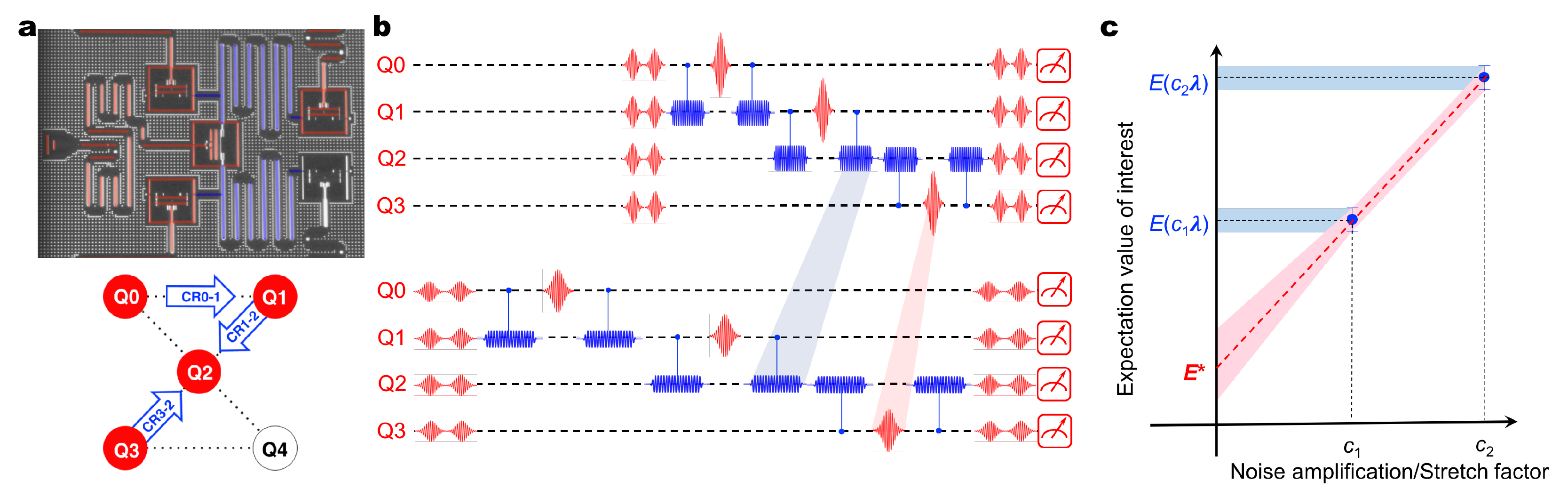}
\caption{\label{Figure3}  \textbf{Device and experimental protocol}  \textbf{a} False-colored optical micrograph (top) of the superconducting quantum processor and schematic (bottom) of the qubits and gates utilized in the experiment. The device is composed of 5 transmon qubits, with the coupling provided by 2 superconducting CPW resonators, in blue. \textbf{b} A measurement of the expectation value after rescaled state preparation is equivalent to a measurement under an amplified noise strength, if the noise is time-translation invariant. \textbf{c} An illustration of the error mitigation method, shown here for a first-order Richardson extrapolation to the zero-noise limit, highlights that the variance of the mitigated estimate $E^*$ is crucially dependent on the variance of the unmitigated measurements, and the stretch factors $c_i$.}
\end{figure*}

We shall first briefly describe the proposal of ~\cite{Mitigation3} and discuss important considerations for our implementation with superconducting qubits. Any quantum circuit can be expressed in terms of evolution under a time-dependent drive Hamiltonian $K(t)=\sum_{\alpha}J_{\alpha}(t)P_{\alpha}$ for a time $T$, where each $P_{\alpha}$ represents a $N$-qubit Pauli operator, and $J_{\alpha}$ is the strength of the associated interaction. The expectation value of an observable of interest $E_K(\lambda)$ for a state prepared by the drive $K$ in the presence of noise can be expressed as a power series around its zero-noise value $E^*$ as 

\begin{align}
E_K(\lambda)=E^*+\sum_{k=1}^{n} a_k\lambda^k+\mathcal{O}(\lambda^{n+1})
\end{align}

Here $\lambda\ll1$ is a small noise parameter, and the coefficients in the expansion $a_k$ are dependent on specific details of the noise model. The primary objective of this paper is to experimentally obtain improved estimates to the noise-less expectation value $E^*$ despite using noisy quantum hardware. 
A powerful numerical technique to suppress the higher order noise terms in Eq. 1 is Richardson's deferred approach to the limit~\cite{Richardson1911}. If $n$ additional estimates to the expectation value $\hat{E}_K(c_i\lambda)$ can be obtained for precisely amplifying the noise rate by factors $c_i$ for $i=1,2..n$, an improved approximation to $E^*$ with a reduced error of order $\mathcal{O}(\lambda^{n+1})$ can be constructed as

\begin{align}
\hat{E}^n_K(\lambda)=\sum_{i=0}^{n}\gamma_i\hat{E}_K(c_i\lambda)
\end{align}

For a chosen set of  ${c_i}$, the coefficients $\gamma_i$ are solutions to $\sum_{i=0}^{n} \gamma_i=1$ and $\sum_{i=0}^{n} \gamma_ic^k_i=0$ for $k=1...n$.  Now, precisely amplifying the strength of the noise is a very challenging experimental task. This can be approximated by the insertion of noisy gates in the quantum circuit, but relies heavily on assumptions of the noise mechanisms at play~\cite{dumitrescu2018}. However, if the noise is time-translation invariant, it can be shown that measurements of the expectation value after evolution under a scaled drive $K^I(t)=\sum_{\alpha}J^i_{\alpha}(t)P_{\alpha}$ for a time $c_iT$ (within the coherence window) is equivalent to a measurement under an amplified noise strength $c_i\lambda$. Beyond this requirement, we emphasize that the method is completely agnostic to the details of the noise model, making it extremely attractive for implementations on near-term noisy hardware. The scaled strength of the interactions in the drive are given as $J^i_{\alpha}(t)=J_{\alpha}(t/c_{i})/c_i$, thereby requiring a good understanding and control of the gates used in the circuit. Also, for superconducting qubits, that often show fluctuations ~\cite{Muller2015} in relaxation $T_1$ and coherence times $T_2$, the requirement of time-translation invariant noise implies that the measurements under the scaled dynamics need to be made within the typical timescales of these fluctuations (see supplementary information). 

The experiments described in this letter are performed on a 5-qubit superconducting processor. The device comprises of fixed-frequency Josephson-junction-based transmon~\cite{Koch2007} qubits, with individual superconducting co-planar waveguide (CPW) resonators for qubit control and readout, and another pair of CPW resonators providing the qubit connectivity. This fixed-frequency architecture is favorable for obtaining long coherence times, and the qubit control and readout is solely by microwave pulses. 

In our device architecture, arbitrary quantum circuits are implemented using combinations of single qubit gates, and two-qubit gates between nearest-neighbor qubits. We shall first discuss the implementation of our error mitigation scheme for single qubit gates in Figure 2(a,b). Single qubit control is achieved using 4$\sigma$ gaussian pulses with a scaled derivative in quadrature to reduce leakage to higher transmon energy levels, and software-implemented Z-gates~\cite{mckay2016efficient}. Each microwave pulse is followed by a buffer time to ensure separation from subsequent pulses. Arbitrary single qubit rotations are constructed using a interleaved sequence of calibrated $X_{\pi/2}$ pulses and Z rotations $U(\vec\theta) = Z_{\theta_1}X_{\pi/2}Z_{\theta_2}X_{\pi/2}Z_{\theta_3}$ where $\vec\theta$ represents the Euler angles. For the error mitigation experiments, the lengths of the 4$\sigma$ Gaussian pulses as well as the buffer times are stretched by the desired stretch factors $c_i$ and calibrated. We first consider sequences of identity-equivalent random single-qubit Clifford operations that return the qubit into the $\ket{0}$ state. We study the decay of the expectation value of the ground state projector ${\langle{\Pi_0}\rangle}$ with increasing length of the Clifford sequence, for the different stretch factors $c_i$, in Figure 2(a). The data is obtained using 10$^5$ samples. Using measurements for each stretch factor $c_i$ and Clifford sequence, we demonstrate suppression of higher order errors in the estimates of ${\langle{\Pi_0}\rangle}$, using up to a third order Richardson extrapolation to the zero-noise limit. Obtaining reasonable error bounds on the extrapolated estimates the encompasses all the sources of error remains a challenge. Although an obvious approach to capture the effect of finite sampling would be to perform a large number of independent experimental runs, this is an extremely time-consuming task. Instead, as detailed in the supplementary information, we employ a bootstrapping technique to simulate the error associated with finite sampling, given an experimental data set. The distributions of numerical outcomes in Figure 2(a) show that higher order extrapolations are increasingly sensitive to the variance of the unmitigated measurements, and highlight the need for a large number of samples. The error mitigation technique may also be visualized by a Bloch sphere picture of a trajectory that begins in the $\ket{0}$ state and in the absence of noise, brings the qubit to its $\ket{1}$ state, along the surface of the sphere. Every $j^{th}$ point along the trajectory is implemented by the unitary operation $U_j$ defined as

\begin{align}
U_{j+1}=Z_{4\theta(j+1)}X_{\theta(j+1)}X_{-\theta(j)}Z_{-4\theta(j)}U_{j}
\end{align}
Here, $\theta(j)=j\pi/30$, the $X$ rotations are implemented as $X_{\theta(j)}=Y_{\pi/2}Z_{\theta(j)}Y_{-\pi/2}$  for $j=$ 0,1..,29 and $U_0$ is the identity gate. We construct the trajectories using measurements of the projectors ${\langle{X}\rangle}$, ${\langle{Y}\rangle}$, and ${\langle{Z}\rangle}$ for each $c_i$ at every point $j$. While the effect of relaxation and dephasing is apparent in the experimental trajectories, the error-mitigated trajectory approaches the final $\ket{1}$ state.

\begin{figure*}
\includegraphics[width=7in]{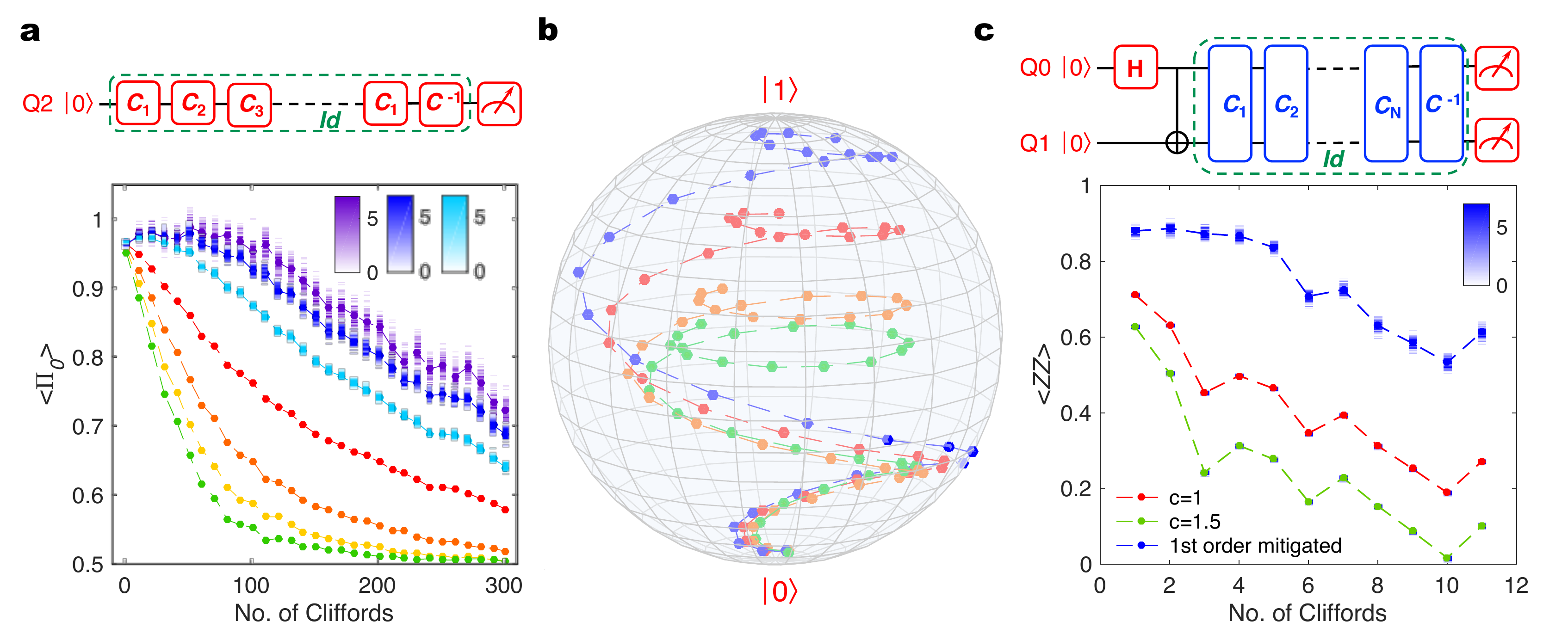}
\caption{\label{Figure2}  \textbf{Error mitigation of random single-qubit and two-qubit circuits}  \textbf{a} Expectation value of the ground state projector for identity equivalent single-qubit Clifford sequences for stretch factors $c=$ 1 (red), 2 (orange), 3 (yellow), 4 (green) and the corresponding Richardson extrapolations to first (light blue), second (dark blue) and third order (violet). \textbf{b} Experimental implementation of trajectories described by Eq. 3, represented on a Bloch sphere for stretch factors $c=$ 1 (red), 2 (orange), 3 (green) and the and the corresponding first-order Richardson extrapolation (blue). The ideal theoretical trajectory is one that takes the qubit from its ground state to its excited state along the surface of the Bloch sphere. \textbf{c} Expectation value of the $ZZ$ parity for identity equivalent two-qubit Clifford sequences applied on a Bell State for stretch factors $c=$ 1 (red), 1.5 (green) and the corresponding 1st order Richardson extrapolations (dark blue). The color density plots of \textbf{a, c} represent histograms of outcomes of 100 numerical experiments obtained by bootstrapping of each experimental data point.}
\end{figure*}

The implementation of this method with two-qubit gates is more challenging for superconducting qubit architectures, since the control is often more complex and the fidelities of two-qubit gates are typically an order of magnitude worse than those of single-qubit gates. Furthermore, the stretching of the gates requires an understanding of the drive Hamiltonian. For our fixed-frequency qubits, we use the all-microwave cross-resonance (CR) gate~\cite{Paraoanu2006,Sheldon2016}, which is implemented by driving a control qubit $Q_c$ with a Gaussian-square-modulated microwave pulse that is resonant with the frequency of the target qubit $Q_t$. For the error mitigation experiments discussed here, we operate our CR gates in a low-power regime where the strengths of the interaction terms in the drive scale linearly with drive amplitude. This is crucial for our implementation since operating the CR gates in a non-linear regime can result in interactions strengths that do not scale appropriately, leading to unphysical values for the mitigated expectation values (see supplementary information). As with single-qubit gates, the pulse lengths, rise-fall times and buffer times of the CR drives are all stretched by the chosen stretch factors and calibrated to a $ZX_{\pi/2}$ gate. As a model experiment that spans the two-qubit Hilbert space, we consider identity-equivalent sequences of random two-qubit Clifford operators that are applied on a maximally entangled Bell state. Figure 2(c) depicts the decay of $ZZ$ parity for these sequences, and its mitigation using a first order Richardson extrapolation to the zero-noise limit. The deviations of the mitigated estimates from the ideal values for the data in Figure 2 can be accounted for by readout assignment errors. It is important to note that the eventual decay of the mitigated curves in Figure 2 shows that the method cannot be applied indefinitely and is ultimately limited by the quantum coherence of the device. Also, gate sequences such as those depicted in Figures 2(a),(c), are often used in standard randomized benchmarking protocols,~\cite{Magesan2012} highlighting the potential applicability of the zero-noise extrapolation technique to novel, improved gate characterization schemes.

\begin{figure*}
\includegraphics[width=7in]{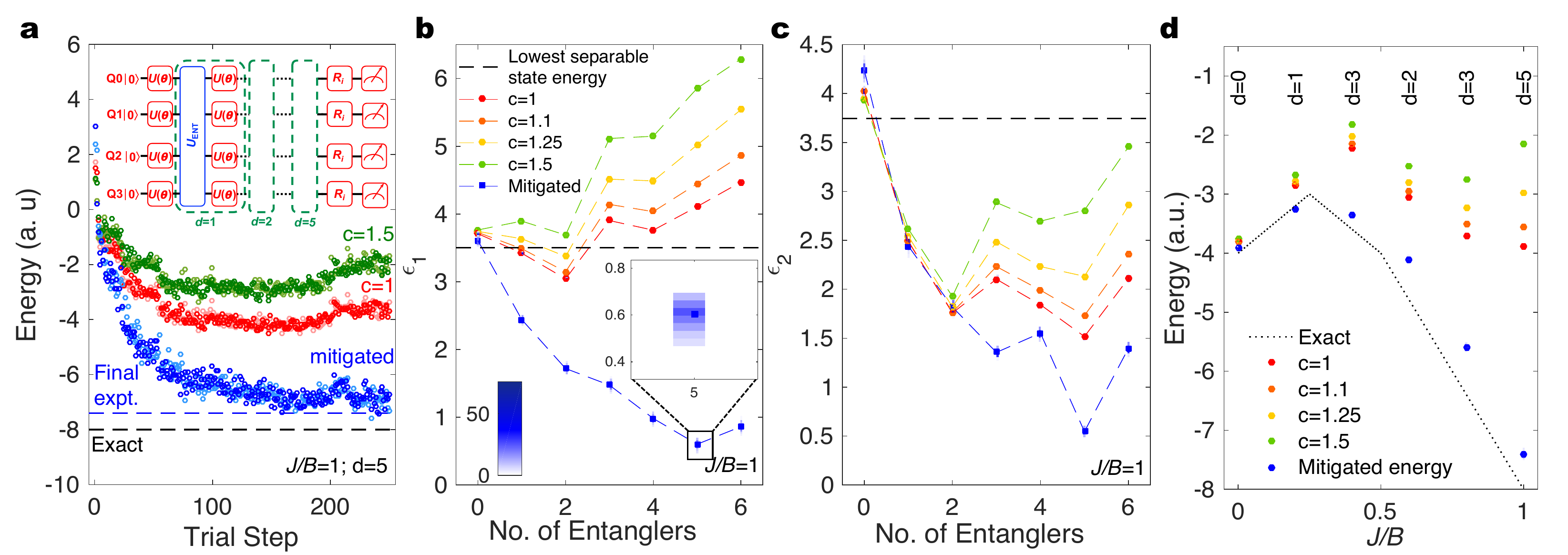}
\caption{\label{Figure2}  \textbf{Application to quantum magnetism}  A hardware-efficient variational eigensolver applied to an anti-ferromagnetic four qubit Heisenberg model in a magnetic field. All color density plots represent histograms of outcomes of 100 numerical experiments obtained by bootstrapping of each experimental data set. \textbf{a} Energy minimization for $J/B=1$ using a depth $d=5$ hardware-efficient trial circuit (depicted by inset) involves the simultaneous optimization of 68 variational parameters. The circuit is constructed as an interleaved sequence of arbitrary single qubit rotations $U(\theta)$ and entanglers $U_\textrm{ENT}$ that entangle all the qubits in the circuit. A set of post-rotations $R_i$ are used to sample the expectation values of the Pauli operators in the target Hamiltonian. At each iteration, the energies of trial states prepared using stretch factors $c=$ 1, 1.5, are measured using 10$^4$ samples, and the mitigated energies obtained by first order Richardson extrapolation are fed to the classical optimization routine. The control parameters from the final 25 iterations are averaged to obtain the controls for the final state measurement. The final energy is then obtained using a linear extrapolation to the energies associated with the final state preparations with stretch factors $c=1,1.1,1.25,1.5$, obtained using 10$^5$ samples. \textbf{b}  Final energy error $\epsilon_1$and \textbf{c} a measure of the error on the individual Pauli terms in the Hamiltonian $\epsilon_2$, as a function of the number of entanglers in the state preparation circuit, for the final stretch factors, and the mitigated values. \textbf{d} Experimental results from the extrapolation of energies obtained from the final stretch factors, compared to the exact energy, for a range of $J/B$ values.}
\end{figure*}

\begin{figure*}
\includegraphics[width=6in]{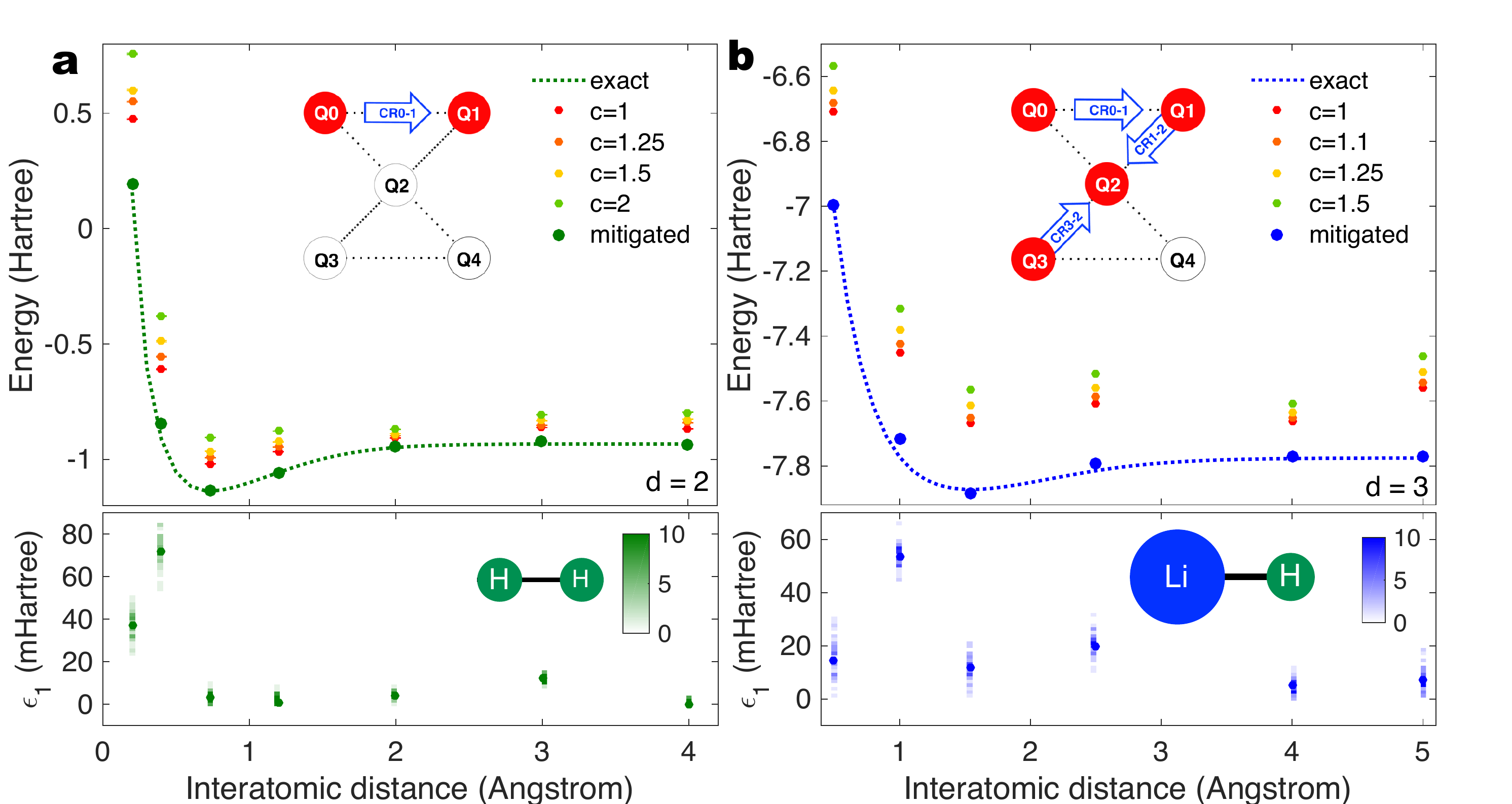}
\caption{\label{Figure4}  \textbf{Application to quantum chemistry}  A hardware-efficient variational eigensolver applied to the electronic structure problem of \textbf{a} H$_2$ and \textbf{b} LiH, using trial state preparation circuit depths $d$= 2 and 3, respectively. The experimental results from the extrapolation of energies obtained from the final stretch factors (see figure legends), compared to the exact energy, for a range of interatomic distances. The bottom panels represent the energy errors $\epsilon_1$ for the Hamiltonians considered with histograms of outcomes of 50 numerical experiments obtained by bootstrapping of each experimental data set. The insets in the top panel depict the qubits used and the gates that compose the entangler $U_\mathrm{ENT}$. The insets in the bottom panel represent schematics of the molecular geometry, not drawn to scale.}
\end{figure*}

The ability to apply the zero-noise extrapolation techniques to random single and two qubit circuits, enables us now to address multi-qubit variational eigensolvers~\cite{Peruzzo13,OMalley16,Kandala2017,Colless2018,Hempel2018arxiv}. Here, variational approximations to the ground state of Hamiltonians of interest are parametrized by experimental controls, and prepared on the quantum processor, with the parameters updated iteratively in conjunction with a classical optimization routine. We first address an interacting spin problem with highly entangled ground states considered in ~\cite{Kandala2017}, specifically, an anti-ferromagnetic four-qubit Heisenberg model on a square lattice, in an external magnetic field: 
\begin{equation}
H=J\sum_{\langle{ij}\rangle}(X_iX_j+Y_iY_j+Z_iZ_j)+B\sum_{i}Z_i
\end{equation}
Here, $J$ is the strength of the spin-spin interaction for nearest neighbor pairs $\langle{ij}\rangle$, and $B$ is the magnetic field in the $Z$ direction. We employ a hardware-efficient variational ansatz ~\cite{Kandala2017}, constructed as an interleaved sequence of arbitrary single qubit rotations and entanglers composed of gates that are natural to the hardware architecture. In the experiment, each entangler is composed of a sequence of echo CR pulses (see supplementary information) and the Euler rotation angles serve as the variational parameters. For the classical optimization routine, we use the simultaneous perturbation stochastic approximation (SPSA) algorithm~\cite{Spall1992} that approximates the gradient at each iteration utilizing only 2 measurements of the cost function, irrespective of the dimensionality of the parameter space. At every iteration, the energies obtained from trial states prepared at stretch factors $c_i=1,1.5$ are used to obtain a mitigated energy estimate which is supplied to the SPSA routine. This is shown for depth $d=5$ trial circuit, using $10^4$ samples along the optimization in Figure 3(a). For the final set of controls, we increase the sampling to $10^5$, and obtain measurements at additional stretch factors $c_i=1,1.1,1.25,1.5$ to reduce the variance on the error-mitigated energy estimate. 

In ~\cite{Kandala2017}, it was numerically shown that the accuracy of a hardware-efficient VQE is affected by the competing effects of circuit depth and incoherent noise. Having discussed the application of zero-noise extrapolation techniques to VQE, we now demonstrate how the suppression of incoherent errors helps discern the improvements with increasing circuit depth.  First, we consider the energy error $\epsilon_1=|E_{\textrm{exp}}-E_{\textrm{exact}}|$ and compare it with the energy error of the lowest energy separable state $\epsilon_{1,d=0}$. The depth dependence of $\epsilon_1$ is depicted in Figure 3(b) for the specific Heisenberg Hamiltonian with $J/B=1$ after optimizations of the kind depicted in Figure 3(a) are run for each $d$. For the $c_i=1$ state preparations, improvements in $\epsilon_1$ with respect to $\epsilon_{1,d=0}$ are limited to $d=2$, due to the decoherence associated with longer trial circuits. However, the ability to suppress the effect of incoherent errors, enables us to discern improvements in $\epsilon_1$ up to $d=5$ trial states. Additionally, we also quantify the errors in the expectation values of the individual Pauli operators $\langle{P_{i }}\rangle$ in the Hamiltonian $H=\sum_{i}\alpha_{i}P_i$ by the quantity $\epsilon_2=\sum_i|\alpha_i|^2(\langle{P_{i,\textrm{exp}}}\rangle-\langle{P_{i,\textrm{exact}}}\rangle)^2$. The improvements in $\epsilon_2$ with increasing circuit depth  shown in Figure 3(c) demonstrates that the error mitigation also improves convergence towards the ground state wave-function, which may not be captured merely by the energy. We map the energy for a range of $J/B$ values in Figure 3(d), algorithmically increasing the circuit depth for every Hamiltonian till no further improvement is obtained in the energy. In similar spirit, one can anticipate the use of such mitigation techniques to also benefit Trotter-based quantum simulations on noisy hardware, enabling improvements in the accuracy with decomposition of the time-evolution into finer steps. 

In contrast to alternate error mitigation schemes~\cite{Mitigation} that rely on specific features of the target Hamiltonian, the zero-noise extrapolation technique is independent of the simulation problem considered. We now demonstrate its general applicability by considering problems in quantum chemistry~\cite{Lanyon2010,Peruzzo13,OMalley16,Kandala2017,Colless2018,Hempel2018arxiv}. in Figure 4. We map the interacting fermion problem for H$_2$ and LiH on two and four qubits respectively, taking advantage of fermionic spin-parity symmetries~\cite{Tapering} and the freezing of core-shell orbitals, as detailed in~\cite{Kandala2017}.  The accuracy of the variational solutions to these problems, obtained on a similar device with comparable coherence times ~\cite{Kandala2017} was severely limited by incoherent errors and insufficient circuit depth for trial state preparations. However, as in the Heisenberg model discussed above, Figure 4 demonstrates far superior accuracies without significant improvements in coherence properties of the hardware, and the ability to benefit from longer circuit depths for trial state preparation. 

A crucial aspect affecting the accuracy and variance of the mitigated estimates are sampling errors. In this context, the integration of fast initialization schemes~\cite{Egger2018} would enable much faster sampling rates. The benefits of this are many-fold: a reduced variance on the mitigated estimates, enabling more accurate classical optimization, reduced experimental run times, the ability to apply higher order Richardson extrapolations, and a reduced effect of coherence time fluctuations in the hardware. 

The work presented here highlights the important considerations for hardware and algorithmic implementations of the zero-noise extrapolation technique, and demonstrates tremendous improvements in the accuracy of variational eigensolvers implemented by a noisy superconducting quantum processor. Further improvements in coherence times will compound to these methods, enabling the applicability to even longer quantum circuits, as well as the ability to reduce the variance of the mitigated estimates with longer stretch factors. Gate calibrations particularly tailored for error mitigation could enable a more accurate rescaling of the dynamics. Alternately, the insertion of twirling gates can be used to randomize systematic errors~\cite{wallman2016} in the rescaling, albeit at the cost of additional measurements. Finally, while the experiments discussed here address problems in quantum simulation, the techniques presented here have great applicability to improved gate characterization, quantum optimization and quantum machine learning~\cite{Havlicek2018} with near-term hardware.

{\bf Supplementary Information} is available in the online version of the paper.

{\bf Acknowledgments}
We thank J. Rozen, M. Takita, and O. Jinka  for experimental contributions, B. Abdo for design and characterization of the Josephson parametric converters, S. Rosenblatt for room temperature characterization, and M. Brink for device fabrication. We acknowledge insightful discussions with S. Brayvi, E. Magesan, S. Sheldon, and M. Takita. The research is based on work supported by the Office of the Director of National Intelligence (ODNI), Intelligence Advanced Research Projects Activity (IARPA), via the Army Research Office contract W911NF-10-1-0324. This research was supported by the IBM Research Frontiers Institute.

{\bf Author contributions}
A.K., K.T., and J.M.G developed the experimental protocol, planned the experiments, and analyzed the experimental data. A.K. performed the experiments. A.D.C. contributed to the experimental set-up. A.M. contributed to the data analysis. A.K., K.T., J.M.C., and J.M.G. wrote the manuscript, with contributions from all authors.

{\bf Author information} Correspondence and request for materials should be addressed to A.K. (akandala@us.ibm.com)

\pagebreak
\onecolumngrid

\beginsupplement

\section*{Supplementary Information: Extending the computational reach of a noisy superconducting quantum processor}

\twocolumngrid

\section{Device and gates}

The quantum processor is composed of five-fixed frequency transmon~\cite{Koch2007s} qubits, and superconducting coplanar waveguide resonators that are employed for qubit-qubit coupling, as well as qubit control and readout, all fabricated on a Si wafer. Each transmon is a single Al-Al$_2$O$_x$-Al Josephson junction, capacitively shunted by Nb capacitor pads. Additional details of the device fabrication may be found in ~\cite{Chow2014s,Corcoles2015s}. The qubit frequencies lie in the $\omega_{01}/2\pi\sim$ 5-5.3 GHz range, and are read out by dispersive measurements through their individual readout resonators at a frequencies close to $\omega_r/2\pi\sim$ 6.5GHz. The qubit anharmonicities are $\sim$ 0.330 GHz. The readout signal is amplified by a Josephson parametric converter (JPC)~\cite{Bergeal2010s,Abdo2011s} at the mixing chamber stage of a dilution refrigerator followed by a high electron mobility transistor at the 4K stage, to achieve typical readout errors  $\epsilon_r<0.05$ for integration times of 2 $\mu$s. Typical relaxation ($T_1$) and dephasing ($T_2$) times of the qubits in the device are in the range 40-70 $\mu$s. 

\begin{figure}
\includegraphics[width=3in]{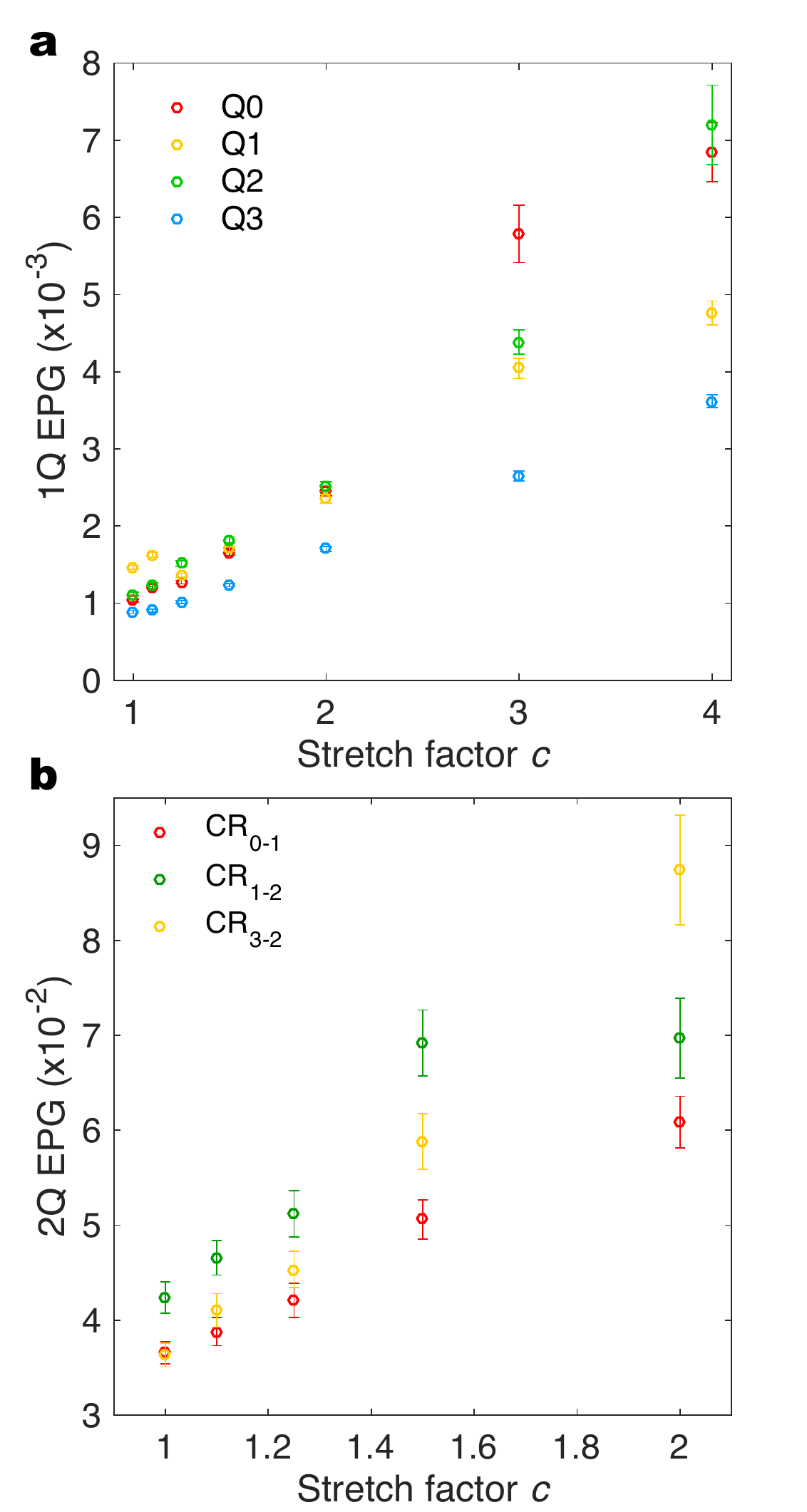}
\caption{\label{RB} {\bf Gate characterization.} {\bf a} Single-qubit and {\bf b} Two-qubit gate fidelities for a range of stretch factors $c_i$ obtained by randomized benchmarking.}
\end{figure}

All single qubit operations are implemented using software $Z$ gates and/or 4$\sigma$ gaussian-modulated microwave pulses employing a derivative removal via adiabatic gate (DRAG) protocol~\cite{Motzoi2009s}. The shortest pulse time (for $c=1$) used in the experiment is  83.3 ns and a buffer time of 6.7 ns is used for ample separation from subsequent pulses. The pulse times, and the buffer times are rescaled by the stretch factor $c_i$, with calibrated pulse amplitudes and DRAG parameters. Figure ~\ref{RB} (a) depicts the gate fidelities of the single qubit gates obtained by simultaneous randomized benchmarking, for a range of different stretch factors. The two-qubit gates are implemented using gaussian-square cross-resonance (CR) pulses~\cite{Chow2011s,Corcoles2013s}. $ZX_{\pi/2}$ operations are constructed using an echo CR (ECR) sequence, detailed in Section III, where a $X_\pi$ gate on the control qubit is sandwiched between two CR pulses of identical pulse time $\tau$ and calibrated amplitude, but opposite sign. The shortest $ZX_{\pi/2}$ gate employed in the experiment employs 2 CR pulses each of width $\tau$= 500 ns, a 3$\sigma$ gaussian rise fall profile (included in $\tau$) with $\sigma=10$ ns, and a buffer time of 6.7 ns. In combination with the $X_\pi$ gate, this brings the total $ZX_{\pi/2}$ gate time for $c=1$ to 1103.4 ns. Note that this is significantly slower than typical operations of the cross-resonance gate~\cite{ibmqx2} -- this is detailed in Section III. As with the single qubit gates, for every stretch factor, the pulse times, $\sigma$, and the buffer times are are rescaled, and the CR pulse amplitudes calibrated to a $ZX_{\pi/2}$ gate. Figure ~\ref{RB} (b) depicts the gate fidelities of the two-qubit gates for a range of stretch factors. 

For a hardware-efficient ansatz, it was shown in reference ~\cite{Kandala2017s} that convergence to the ground state of small molecular and spin systems can be obtained for a range of entangling gate phases around points of maximum concurrence. For the variational trial states implemented in this work, the entangler is constructed as a series of pair-wise $ZX_{\pi/4}$ gates, each employing a ECR pulse sequence. These entanglers are so chosen  to deliver sufficient entanglement, while reducing the effect of decoherence during state preparation.

\section{Coherence time fluctuations in superconducting qubits}

For the zero-noise extrapolation scheme discussed in this work, the measurement of an expectation values under an amplified noise strength is equivalent to a measurement after a rescaled state preparation only under the assumption that the noise is time-translation invariant. This is important to note for implementations of this scheme with superconducting  qubits which show fluctuations in their coherence time. Figure S2(a) depicts fluctuations in the $T_1$ and $T_{2}$ times over a period of approximately 2 hours that are clearly larger than the error bars of the exponential fits to the decay from individual runs. It is therefore important that measurements for different stretch factors that are used to perform Richardson extrapolation for the same quantity are performed shortly after one another.

Over the entire duration of averaging the measurements, the average decay time for the different stretched experiments is then the same, as seen in Figures S2 (b) and (c) for typical $T_1$ and echo $T_{2}$ sequences for $c_i=1,2$ obtained with $10^5$ samples. Alternately, when the different stretched experiments are performed separately, with larger time intervals between them, the decays may differ for different $c_i$, potentially leading to Richardson extrapolated expectation values that are out of bounds. The experiments discussed here employed a sampling rate of 2 kHz, set by the time required for the qubits to naturally relax to the ground state (~5-10 $T_1$) for initialization. The integration with fast initialization schemes will improve the accuracy of the mitigated estimates even further by allowing a larger number of samples in a shorter time, while reducing the susceptibility to coherence fluctuations.

\begin{figure*}
\includegraphics[width=6.5in]{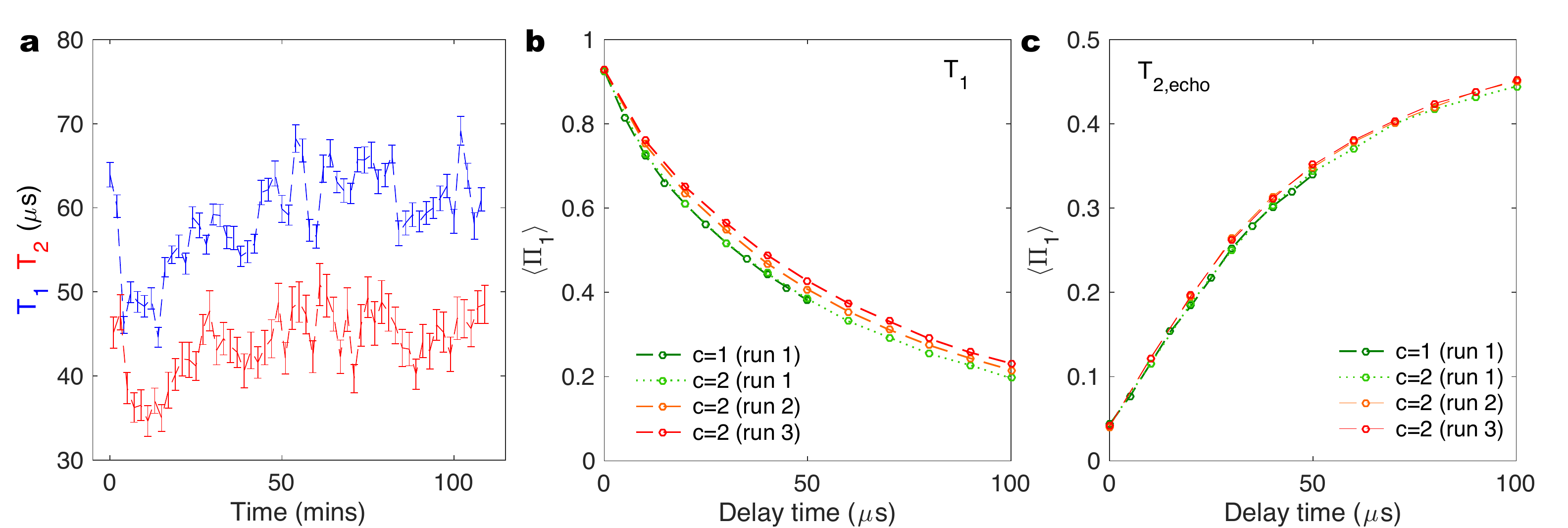}
\caption{\label{Circuit} {\bf Coherence fluctuations in superconducting qubits} {\bf a} Repeated measurements of $T_1$ and $T_2$ show fluctuations that are larger than the error bars to the fits. Decay of excited state projector $\langle{\Pi_1}\rangle$ for a standard {\bf b} $T_1$ sequence and  {\bf c} $T_2$ echo sequence, for stretch factors $c_i$=1,2 obtained using 10$^5$ samples. In both cases, the average decay over the total sampling duration for the different stretch factors is the same, when their measurements are grouped together (run 1), in contrast to when sampled separately (run 2, run 3).}
\end{figure*}

\section{Stretching of two-qubit gates}

We employ the cross resonance (CR) interaction for our two-qubit entangling gate~\cite{Paraoanu2006s,Rigetti2010s,Chow2011s}. This is particularly well suited for our fixed-frequency, all microwave control hardware architecture, and is implemented by driving a control qubit with a microwave tone that is resonant with a nearest neighbor target qubit. Advances in  the understanding of the CR drive hamiltonian have led to controlled-NOT (CNOT) gate fidelities exceeding 99$\%$, with gate times less than 200 ns ~\cite{Sheldon2016s}. An effective Hamiltonian model of the CR drive ~\cite{Magesan2018s} that accounts for higher energy levels of the transmon reveals the following interactions: $IX, IY, IZ, ZX, ZY, ZZ$ and a Stark shift $ZI$ arising due to the non-resonant tone. The $ZX$ and $IX$ terms are predicted to be the dominant interactions, and $ZZ$ and $IZ$ are small and independent of the strength of the CR drive. This is also seen experimentally, in Fig. S3, which depicts the strengths of the different interactions for a range of drive amplitudes, obtained by tomography of the CR drive hamiltonian for the Q3 (control) - Q2 (target) pair. Isolating the $ZX$ interaction enables the construction of a CNOT gate, merely requiring additional single qubit gates. This can be achieved by using a standard echo CR (ECR) sequence, that applies a $X_\pi$ on the control qubit between CR pulses of opposite sign, to refocus the $IX, ZZ,$ and $ZI$ terms. In the absence of classical cross-talk, the $IY$ is negligible, else a cancellation may be employed as in reference ~\cite{Sheldon2016s}. Finally, the phase of the CR drive may be set such that the sole conditional interaction is $ZX$.

\begin{figure}
\includegraphics[width=3.5in]{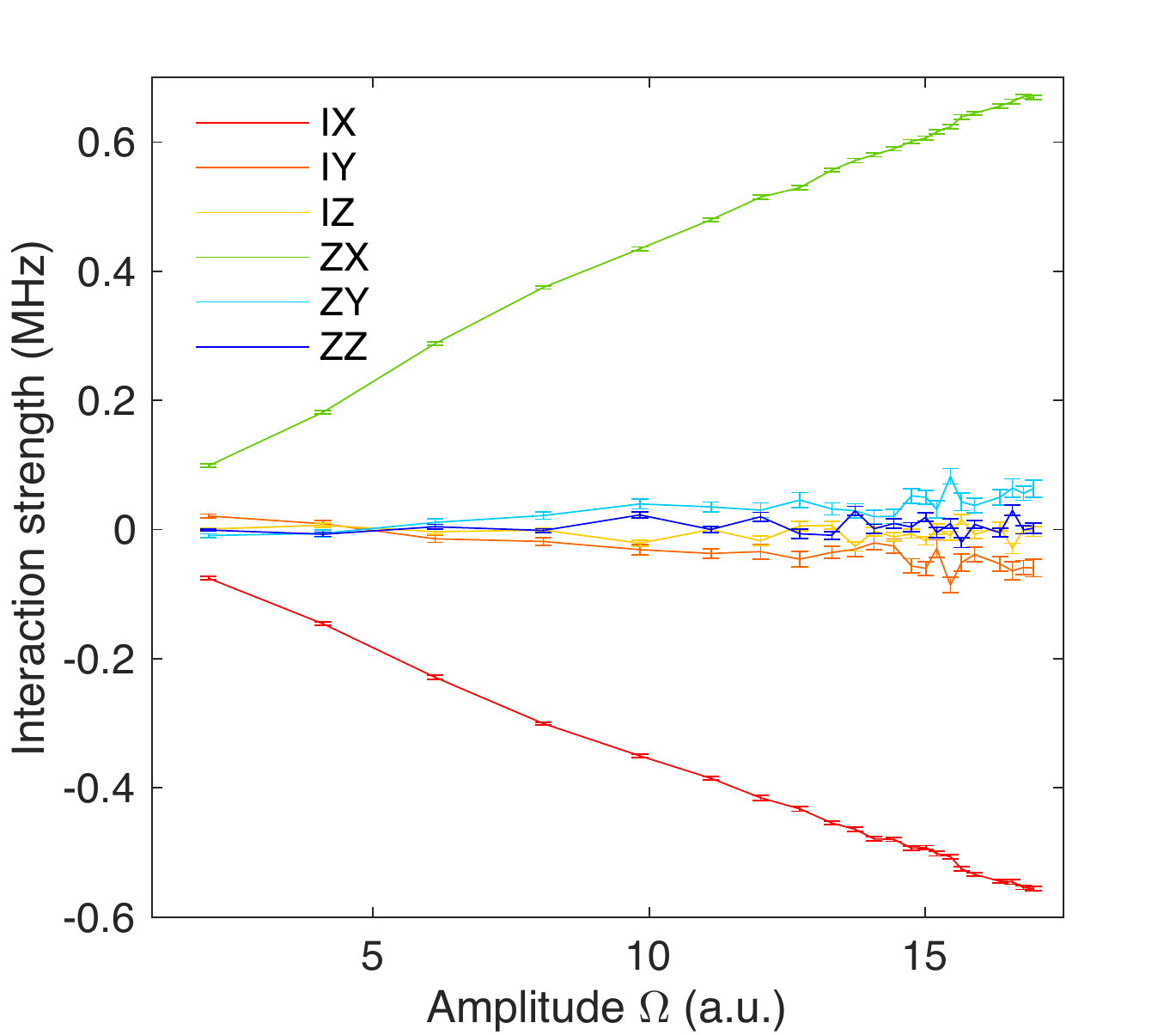}
\caption{\label{Circuit} {\bf Hamiltonian tomography of the cross-resonance drive} Amplitude dependence of the interaction terms in the cross-resonance drive CR$_{3-2}$. The dominant interactions in the drive are $ZX$ and $IX$, with discernible non-linearities in the strong drive limit.}
\end{figure}

\begin{figure*}
\includegraphics[width=7in]{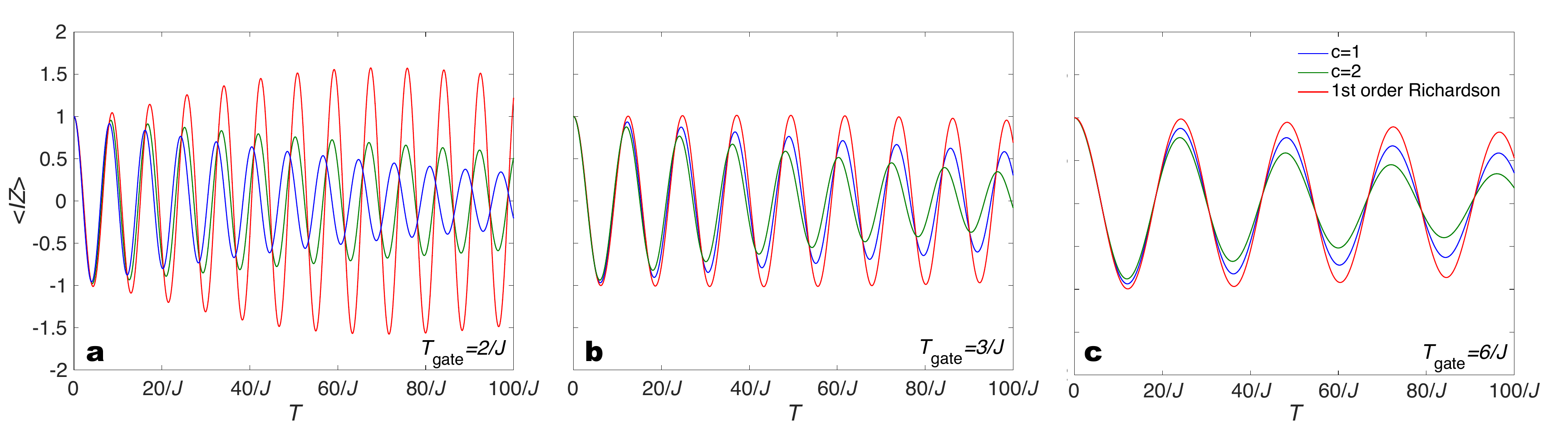}
\caption{\label{Circuit} {\bf Error mitigation with a non-linear cross resonance drive } Time evolution of $\langle{IZ}\rangle$ under a simplistic CR drive $J_{ZX}=-0.0159J\Omega+1.0541\times10^{-6}J\Omega^{3}$, and fixed amplitude damping and dephasing channels, for drive amplitudes calibrated to $ZX_\pi/2$ gate times $T_{\textrm{gate}}=$ {\bf a} $2J^{-1}$, {\bf b} $3J^{-1}$, {\bf c} $6J^{-1}$ and stretch factors $c=$1 (blue), 2 (green). The mitigated time evolution (red) goes out of bounds for fast gates (large drive amplitudes), due to the non-linearity in the amplitude dependence.}
\end{figure*}

As seen in Figure S3, the amplitude dependence of the interactions in the CR drive can depict small non-linearities that are important to consider for the implementation of our error mitigation scheme. These non-linearities have been detailed in the recent work of ~\cite{Magesan2018s}. For instance, the amplitude dependence of the $ZX$ interaction, obtained by a perturbative model to third order, takes the following form

\begin{equation}
\begin{aligned}
J_{ZX}= & -\Omega\Bigg(\frac{J\delta_1}{\Delta(\delta_1+\Delta)}\Bigg)+ \\ &\Omega^3\Bigg(\frac{J\delta_1^2(3\delta_1^3+11\delta_1^2\Delta+15\delta_1\Delta^2+9\Delta^3)}{4\Delta^3(\delta_1+\Delta)^3(\delta_1+2\Delta)(3\delta_1+2\Delta)}\Bigg).
\end{aligned}
\end{equation}

Here, $\Delta$ is the frequency difference between the qubits, $\delta_1$ represents their anharmonicity, $J$ is the qubit-qubit coupling and $\Omega$ is the amplitude of the CR drive. If the gates are not tuned-up appropriately this non-linearity can lead the stretched evolution to be out of phase. To model the effect of this non-linearity on our error mitigation scheme , we use a simplistic form of the CR drive $J_{ZX}=-0.0159 J \Omega+1.0541\times10^{-6} J \Omega^{3}$, obtained from Eq. 1 using $\delta_1$=320 MHz and $\Delta$=50 MHz and add fixed strength $\lambda = 2\times10^{-3} J$ amplitude damping and de-phasing noise for each qubit. The simplistic CR model system evolves as
\bq
\partial_t \rho &=& -i J_{ZX}[ZX,\rho]  \no
&+&\lambda \sum_{i=1}^2\left(\sigma_i^- \rho \sigma_i^+ - \frac{1}{2}\{\sigma_i^+ \sigma_i^-,\rho\} + Z_i\rho Z_i - \rho \right)
\eq
for a total time $T = 100 J^{-1}$ with $\sigma^{\pm} = 2^{-1/2}(X \pm i Y)$. In Figure S4 we compare the decay of oscillations in the $\langle{IZ}\rangle$ for 3 different drive amplitudes of this model. These were chosen to implement a $ZX_{\pi/2}$ gate with three different gate times $T^a_{\textrm{gate}} = 2J^{-1}$, $T^b_{\textrm{gate}} = 3 J^{-1}$ and $T^c_{\textrm{gate}} = 6 J^{-1}$ in Figure S4 (a),(b) and (c) respectively. We used the relationship $\Omega = \pi\Delta(\delta_1+\Delta)(2 T_{\textrm{gate}}J\delta_1)^{-1}$ to fix the drive amplitudes. The drive time and amplitudes are scaled for the stretch factors $c_i=1,2$ and the mitigated estimates are then obtained using  a first-order Richardson extrapolation. Figure S4(a) shows that the large non-linearity for fast gates drives the mitigated estimates severely out of bounds. This may be visualized as a consequence of the "out-of-phase" oscillations in $\langle{IZ}\rangle$ for the two stretch factors. However, the effect of the non-linearity may be suppressed by using a weaker drive amplitude, as seen in Figure S4(c), where the extrapolation is well within bounds, although at the expense of slower gate times. In the absence of the non-linear term in the model, it can be seen that the mitigated estimates remain in bounds even at high drive powers.

Typical operations of high fidelity cross resonance gates are often in a fast, non-linear regime. This highlights the need for gate calibrations specifically tailored for error mitigation, since a simple rescaling of the amplitude would not suffice, as shown above. In the experiment, the CR pulse amplitudes are calibrated to a $ZX_{\pi/2}$, for each stretch factor $c_i$. In order to reduce the effect of non-linearities in the other terms in the drive Hamiltonian, we employ the echo sequence described above to isolate the $ZX$ interaction, as well as conservatively operate the gate in a more linear, low power regime. The two-qubit gate times are therefore significantly slower than normal operation. However, the accuracies of the computations achieved in this work with these slow gates are otherwise inaccessible with faster gates in the absence of error mitigation.  It is also important to note that the the error - mitigation protocol cancels error terms that are invariant under rescaling~\cite{Mitigation3}.  This also includes constant terms in the Hamiltonian evolution such as the static $ZZ$ - interaction.

\section{Sampling and Bootstrapping}

In addition to the effect of finite sampling, the accuracy of measured expectation values is sensitive to the readout assignment infidelity, in particular for large weight Pauli operators. For the variational eigensolver described in Figures 3 and 4 of the main text, the expectation values are therefore corrected using a readout calibration of the possible outcomes, at every iteration, as in ~\cite{Kandala2017s}. The variance of the measured expectation values for the different stretch factors then translates to error bounds on the mitigated estimates obtained from a zero-noise extrapolation. One obvious way to estimate the effect of finite sampling on the error bounds of the mitigated estimates is to simply repeat the the experiment a large number of times. However, since this is an extremely time consuming task, we simulate the spread arising from finite sampling, using a well-known statistical technique, bootstrapping~\cite{Efron1992s}. Using the experimentally measured probability distributions, we resample both, the readout calibrations, as well as the measurements of the quantum state of interest. The resampled probability distributions are then used to evaluate the assignment-error-corrected expectation values, for each stretch factor, and consequently, the mitigated estimates. Running this bootstrapping protocol 50-100 times is then used to obtain a distribution of numerical outcomes for the mitigated expectation value.

\end{document}